\documentclass[aps,prl,twocolumn,superscriptaddress,showpacs]{revtex4}
\usepackage{amssymb}
\usepackage{graphicx}
\usepackage[tbtags]{amsmath}

\begin{document}
\title{Explosive or Continuous: Incoherent state determines the route to synchronization}
\author{Can Xu}
\affiliation{Department of Physics and the Beijing-Hong Kong-Singapore Joint Centre for Nonlinear and Complex Systems (Beijing), Beijing Normal University, Beijing 100875, China}
\author{Jian Gao}
\affiliation{Department of Physics and the Beijing-Hong Kong-Singapore Joint Centre for Nonlinear and Complex Systems (Beijing), Beijing Normal University, Beijing 100875, China}
\author{Yuting Sun}
\affiliation{Department of Physics and the Beijing-Hong Kong-Singapore Joint Centre for Nonlinear and Complex Systems (Beijing), Beijing Normal University, Beijing 100875, China}
\author{Xia Huang }
\affiliation{Department of Mathematics and Physics, North China Electric Power University,Beijing 102206, China}
\author{Zhigang Zheng}
\email[]{zgzheng@bnu.edu.cn}
\affiliation{Department of Physics and the Beijing-Hong Kong-Singapore Joint Centre for Nonlinear and Complex Systems (Beijing), Beijing Normal University, Beijing 100875, China}
%\date{\today}

\begin{abstract}
Collective behaviors of coupled oscillators have attracted much attention. In this Letter, we propose an ensemble order parameter(EOP) equation that enables us to grasp the essential low-dimensional dynamical mechanism of the explosive synchronization in heterogeneous networks. Different solutions of the EOP equation build correspondences with diverse collective states, and different bifurcations reveal various transitions among these collective states. The structural relationship between the incoherent state and synchronous state leads to different routes of transitions to synchronization, either continuous or discontinuous. The explosive synchronization is determined by the bistable state where the measure of each state and the critical points are obtained analytically by using the EOP equation. Our method and results hold for heterogeneous networks with star graph motifs such as scale-free networks, and hence, provide an effective approach in understanding the routes to synchronization in more general complex networks.
\end{abstract}


\maketitle

Understanding the intrinsic microscopic mechanism of collective behavior of populations of coupled units has become a focus in a variety of fields, such as biological neurons circadian rhythm, chemically reacting cells, and even society systems \cite{1,2,3,4,5,6,6'}. In particular, when the frequencies of nodes are positively correlated to the node's degrees, an abrupt transition from incoherent state to synchronization in heterogenous networks takes place \cite{7}. This explosive synchronization(ES) has been observed in scale-free(SF) networks, and electronic circuits \cite{10}. Numerous efforts have been made to understand ES from different viewpoints such as the topological structures of networks and coupling functions among nodes \cite{8,9,10,11,12,13,13'}. The key point in understanding the discontinuous synchronization transition is the analysis of the multistability and competition of miscellaneous synchronous attractors in phase space, for example the hysteretic behavior at the onset of synchronization \cite{19}. However it is difficult to get an analytical insight in a high-dimensional phase space, and a convincing understanding is still lacking.

It is our motivation in this Letter to reveal the mechanism of synchronization transition, especially ES in networks with a star motif by analyzing in a low-dimensional complex ensemble order parameter(EOP) space in terms of the Ott-Antonsen method. Different solutions of the EOP equation build correspondences with diverse collective states, and different bifurcations reveal various transitions among these collective states. The structural relationship between the incoherent state and synchronous state leads to different routes of transitions to synchronization, either continuous or discontinuous. ES is a touchstone in tesifying our approach. We reveal that ES is attributed to the coexistence of the incoherent state and the attractive synchronous state. The explosive synchronization is determined by the bistable state where the measure of each state and the critical points are obtained analytically by using the EOP equation. The scenario is further applied to discussions of ES in generic SF networks.

In a heterogeneous network, such as a SF network, hubs play a dominant role. Hence a star motif with a central hub is a typical topology in grasping the essential property of the heterogeneous networks. By keeping oscillators on $K$ leaf nodes with the same frequency $\omega$ and the hub with $\omega_{h}$, the equations of motion can be written as
\begin{equation}\label{equ:0}
\begin{aligned}
\dot{\theta}_{h} &=\small{\omega_{h}+\lambda\, \sum_{j=1}^{K}\sin\,(\theta_{j}-\theta_{h})},\\
\dot{\theta}_{j} &=\small{\omega+\lambda\,\sin\,(\theta_{h}-\theta_{j})},
\end{aligned}
\end{equation}
where $\small{1\leq j\leq K}$, $\theta_{h},\theta_{j}$ are phases of the hub and leaf nodes, respectively, $\lambda$ is the coupling strength. By introducing the phase difference $\varphi_{j}=\theta_{h}-\theta_{j}$, Eqs. (\ref{equ:0}) can be transformed to
\begin{equation}\label{equ:15}
\dot\varphi_{j}=\Delta\omega-\lambda\,\small{\sum_{i=1}^{K}}\,\sin(\varphi_{i})-\lambda\,\sin\varphi_{j},
\end{equation}
where $\Delta\omega=\omega_{h}-\omega$ is the frequency difference between the hub and leaf nodes.

Synchronous state(SS) is defined as $\varphi_{i}(t)=\varphi_{j}(t)\equiv \varphi(t)$ and $\dot\varphi(t)=0$, which can be solved as $\sin\varphi=\Delta\omega/(K+1)\lambda$ from Eqs. (\ref{equ:15}). Since $|\sin\varphi|\leq 1$, the synchronization state exists when $\lambda\geq\lambda_{c}=\Delta\omega/(K+1)$. The synchronous state is found to be stable when $\lambda\geq\lambda_{c}$ by using linear-stability analysis. Further numerical computations reveal that the transition to SS is abrupt, and there is a hysteretic behavior at the onset of synchronization. $\lambda_{c}^{b}$ and $\lambda_{c}^{f}$ are the backward and forward critical coupling strengths respectively, where $\lambda_{c}^{b}=\lambda_{c}$ and  $\lambda_{c}^{f}$ depends on initial states. The upper limit of $\lambda_{c}^{f}$ is denoted by $\hat\lambda_{c}^{f}$. As $\lambda>\hat\lambda_{c}^{f}$, SS is globally attractive. The region between $\lambda_{c}^{b}$ and $\hat\lambda_{c}^{f}$ is the coexistence regime for SS and incoherent state. The dynamic process of the synchronization transition depends crucially on the basin of attraction of each state \cite{19}. However, it is hard to investigate the intermingling structure of these different attractors in the $K$-dimensional phase space \{$\phi_{i}, i=1,2,...,K$\}, and till now only numerical works for not large $K$ have been done. It is significant to find an analytical scheme to excavate the coexistence of synchronous and incoherent attractors and quantitatively reveal the mechanism of ES.

By introducing the order parameter $z=re^{i\Phi}=\frac{1}{K}\sum_{j=1}^{K}e^{i\varphi_{j}}$, it is instructive to rewrite Eqs. (\ref{equ:15}) as
\begin{equation}\label{equ:2}
\dot\varphi_{j}=fe^{i\varphi_{j}}+g+\bar{f}e^{-i\varphi_{j}},\quad j=1,\cdots,K,
\end{equation}
where $i$ denotes the imaginary unit and $f=i\frac{\lambda}{2}$, $g=\Delta\omega-\lambda Kr\sin(\Phi)$. In terms of Watanabe-Strogatz's approach, the phase dynamics of $K$ nodes can be constructed from $K$ constants \{$\xi_{j},1\leq j\leq K$\} as $e^{i\varphi_{j}(t)}=M_{t}(e^{i\xi_{j}})$, where $M_{t}(x)=\frac{e^{i\phi(t)}x+\beta(t)}{1+\bar\beta(t)e^{i\phi(t)}x}$ is a Mobius transformation \cite{18,22}. By applying it to Eqs. (\ref{equ:2}) one obtains
\begin{equation}\label{equ:25}
\begin{aligned}
\dot\beta &=i(f\beta^{2}+g\beta+\bar f),\\
\dot\phi &=f\beta+g+\bar f\bar\beta.
\end{aligned}\end{equation}

For the situation of thermodynamic limit $K\rightarrow\infty$ and the uniform measure of phases, the evolution of $\beta(t)$ and $\phi(t)$ in Eqs. (\ref{equ:25}) can be separated. We thus get $\beta(t)=z(t)$ and the equation of the order parameter as $\dot z=i(fz^{2}+gz+\bar f)$. For a finite $K$, we can get the measure and the distribution of phases and further define an ensemble order parameter (EOP) as $\small{z=\left \langle\sum_{j=1}^{K}e^{i\varphi_{j}}/K\right \rangle}$ through an ensemble consisting of systems with same parameter and random initial conditions confined in an interval $[\theta_{a},\theta_{b}]$. The equation of EOP of Eqs. (\ref{equ:2}) is
\begin{equation}\label{equ:26}
\dot z=-\dfrac{\lambda}{2}z^{2}+i(\Delta \omega-\lambda Kr\sin\Phi)z+\dfrac{\lambda}{2}.
\end{equation}
Eq. (\ref{equ:26}) describes the collective dynamics of Eqs. (\ref{equ:15}) in terms of the order parameter. In the phase space of EOP, the synchronous state corresponds to a fixed point with $r=1$ and a fixed phase $\Phi$. All the other solutions of Eq. (\ref{equ:26}) represent various incoherent states. Some typical incoherent states include the splay state defined by $r<1$ with a fixed phase $\Phi$, the in-phase state defined by $r=1$ with a periodic phase $\Phi(t)$ and the neutral state defined by time-periodic $r(t)$ and $\Phi(t)$. The transitions from these states to synchronization correspond to different scenarios of collective behaviors.

Eq. (\ref{equ:26}) can be rewritten in cartesian coordinates $z=x+iy$ as
\begin{equation}\label{equ:3}
\begin{aligned}
\dot x &=\lambda(K+\frac{1}{2})y^{2}-\frac{\lambda}{2}x^{2}-\Delta\omega y +\frac{\lambda}{2},\\
\dot y &=-\lambda(K+1)xy+\Delta\omega x.
\end{aligned}
\end{equation}
Eqs. (\ref{equ:3}) is invariant under the time-reversal transformation $(t,x)\rightarrow(-t,-x)$. This implies the quasi-Hamiltonian property of Eqs. (\ref{equ:3}) \cite{23}, where the phase volume in the vicinity of any periodic orbits is conserved.

\begin{figure}
  \includegraphics[width=7.5cm,height=4.8cm]{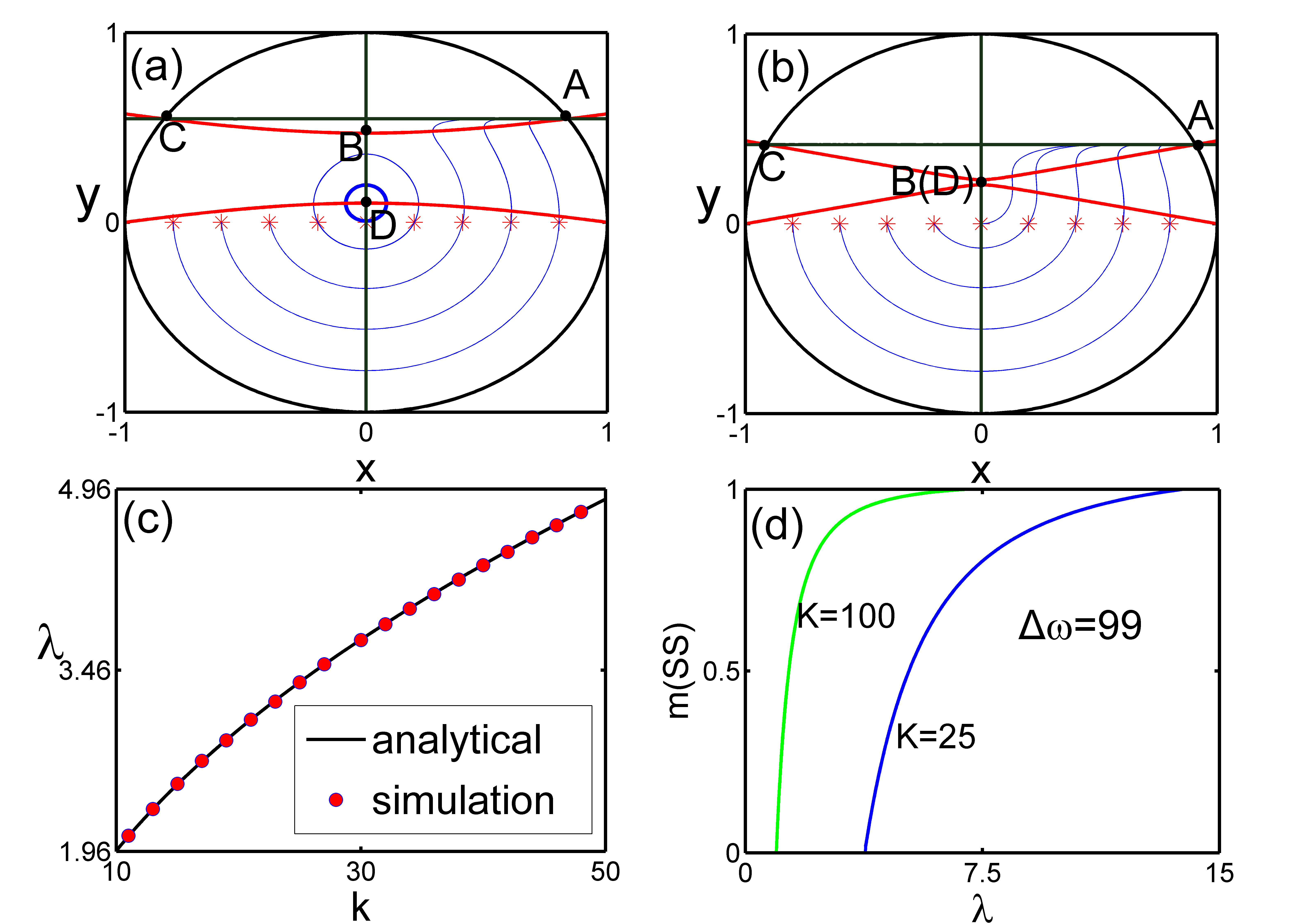}\\
  \caption{(color online) EOP phase plane of Eqs. (\ref{equ:3}) with $\Delta\omega=9$, $K=10$, (a) $\lambda=1.5$, (b) $\lambda=1.9$. Red lines are $\dot x=0$, and black $\dot y=0$. The intersections of them are fixed points A-D. Different initial values with trajectories are marked by '$\ast$'. (c) Upper limit of forward critical coupling strength in Eqs. (\ref{equ:0}). (d) The measure of SS against the coupling strength for different $K$.}\label{fig:1}
\end{figure}

In the phase space of the EOP, the natural boundary of order parameter is $x^{2}+y^{2}=1$. A fixed point is determined by the intersection of nullclines  $\dot x=0$ and $\dot y=0$ within the boundary. When the coupling $\lambda$ is small enough, there is only one fixed point, which is neither an attractor nor a repeller for they should appear in pairs. Hence incoherent states are neutrally stable periodic orbits around the fixed point. This can be verified by using linear stability analysis.

In the bistable regime, as shown in Fig. \ref{fig:1}({a), the nullclines $\dot x=0$ and $\dot y=0$ have four intersections labeled by A-D with A an attractor, C a repeller and B, D neutrally stable. Any orbits crossing the nullcline A-B-C will eventually fall to A, and others will hold the property as periodic orbits. It is clear that the stable fixed point A corresponds to the synchronous state(SS).

As $\lambda$ increases, points D and B close to each other and eventually collide at a critical coupling, as shown in Fig. \ref{fig:1}(b), and the SS becomes globally attractive. This critical coupling corresponds to the upper limit of $\lambda_{c}^{f}$, which can be determined as
\begin{equation}\label{equ:35}
\small{\hat\lambda_{c}^{f}=\Big(\dfrac{\Delta\omega}{\sqrt{K}}\dfrac{1}{\sqrt{2+K^{-1}}}\Big)}.
\end{equation}
The analytical curve and numerical results are given in Fig. \ref{fig:1}(c). An approximation of this result was previously estimated as $\hat\lambda_{c}^{f}\approx 0.6989\Delta\omega/\sqrt{K}$ \cite{19}, but this estimation is only the limiting case of Eq. (\ref{equ:35}) for large $K$ where $\lambda_{c}^{f}=\Delta \omega /\sqrt{2K}$.

It is important to compute the measure of SS in phase space when SS and the incoherent state coexist. The measure is defined as is $m(\textmd{SS})=S_{syn}/S_{0}$, where $S_{0}$ and  $S_{syn}$ are the volume of the whole phase space and the volume of the basion of attraction of SS respectively. This can be analytically obtained in the EOP space as $S_{syn}\approx\pi(1-(x_{\textmd{B}}-x_{\textmd{D}})^{2}-(y_{\textmd{B}}-y_{\textmd{D}})^{2})$ and $S_{0}=\pi$, where $(x_{\textmd{B}},y_{\textmd{B}})$ and $(x_{\textmd{D}},y_{\textmd{D}})$ are the coordinates for points B and D respectively. Therefore the measure of SS is
\begin{equation}
\footnotesize{m(\textmd{SS})=1+\dfrac{4}{(1+2K)}-\dfrac{4\Delta\omega^{2}}{\lambda^{2}(1+2K)^{2}}+\delta_{\lambda}, \lambda_{c}^{b}\leq\lambda\leq\hat\lambda_{c}^{f}},
\end{equation}
where $\delta_{\lambda}$ is the correction factor $\delta_{\lambda}\approx2(\lambda-\hat\lambda_{c}^{f})^{2}/\Delta\omega^{2}$ bounded by $1/K$. When $\lambda\leq\lambda_{c}^{b}$, $m(\textmd{SS})=0$ and when $\lambda\geq\hat\lambda_{c}^{f}$, $m(\textmd{SS})=1$. In Fig. \ref{fig:1}(d), $m(\textmd{SS})$ vs $\lambda$ is shown. When $K\rightarrow\infty$, the measure can be approximated by $m(\textmd{SS})\approx1-\lambda^{-2}$.

The above results indicate that the EOP approach can successfully describe the collective dynamics of coupled oscillators, and the EOP equation provides an exact description in revealing the transitions, coexistence and competitions between incoherent and synchronous states. However, the quasi-Hamiltonian property of system Eqs. (\ref{equ:3}) should be a specific case depending crucially on the coupling function. Therefore we consider more general cases by adopting the Kuramoto model Eqs. (\ref{equ:0}) with a phase shift:
\begin{equation}\label{equ:4}
\begin{aligned}
\dot{\theta}_{h} &=\omega_{h}+\lambda\, \sum_{j=1}^{K}\sin\,(\theta_{j}-\theta_{h}-\alpha),\\
\dot{\theta}_{j} &=\omega+\lambda\,\sin\,(\theta_{h}-\theta_{j}-\alpha),
\end{aligned}\end{equation}
where $-\pi/2\leq\alpha\leq \pi/2$ is the phase shift, with $\alpha=0$ corresponding to the case of Eqs. (\ref{equ:0}). When $\alpha=0, \pm\frac{\pi}{2}$ the equation is time reversible, and they divide the parameter space $\alpha$ into two dynamical regimes $(-\pi/2,0)$ and $(0,\pi/2)$.

By introducing $\varphi_{j}=\theta_{h}-\theta_{j}$, Eqs. (\ref{equ:4}) are transformed to $\dot\varphi_{j}=\Delta\omega-\lambda\,\sum_{i=1}^{K}\,\sin(\varphi_{j}+\alpha)-\lambda\,\sin(\varphi_{j}-\alpha)$, which is described by the following EOP equation
\begin{equation}\label{equ:5}
\dot z=-\dfrac{\lambda}{2}e^{-i\alpha}z^{2}+i(\Delta \omega-\lambda Kr\sin(\Phi+\alpha))z+\dfrac{\lambda}{2}e^{i\alpha}.
\end{equation}

The condition for existence of the SS can be obtained similarly as
\begin{equation}
\small{\lambda\geq\lambda_{ec}=\dfrac{\Delta\omega}{\sqrt{(K-1)^{2}+4K\cos^{2}\alpha}}}.
\end{equation}
Further linear-stability analysis of the EOP equation Eq. (\ref{equ:5}) indicates that the SS is stable when
\begin{equation}\begin{cases}\begin{array}{ll}
\lambda\geq\lambda_{sc}^{-}=\footnotesize{\dfrac{\Delta\omega}{K\cos2\alpha+1}}, & \footnotesize{for\, \alpha \in(\alpha_{0}^{-}, 0]}, \\
\lambda\leq\lambda_{sc}^{+}=\footnotesize{\dfrac{-\Delta\omega}{K\cos2\alpha+1}}, & \footnotesize{for\, \alpha \in(\alpha_{0}^{+}, \pi/2 ]},
\end{array}\end{cases}\end{equation}
where $\alpha_{0}^{\pm}=\pm\arccos(-1/K)/2$. When $-\pi/2\leq\alpha\leq\alpha_{0}^{-}$ the SS is always unstable, and when $0\leq\alpha\leq\alpha_{0}^{+}$ the SS is always stable.

When $\lambda$ is small enough, Eq. (\ref{equ:5}) always has one fixed point within the boundary in phase space $(x,y)$. When $\alpha=0, \pm\pi/2$, the fixed point is neutrally stable. However when $\alpha\in(-\frac{\pi}{2},0)$, this point is a stable attractor, which is physically a splay state(SPS), when $\alpha\in(0,\frac{\pi}{2})$, this point is an unstable repeller, and all orbits will evolve to the boundary as a limit circle that physically corresponds to the in-phase state(IPS). For Eqs. (\ref{equ:4}), SPS is the state where phase differences between hub and leaf nodes satisfy $\varphi_{i}(t)=\varphi(t+\frac{iT}{K})$ with $T$ the period of $\varphi(t)$, and IPS is the state with $\varphi_{i}(t)=\varphi(t)$. Different routes of transitions from SPS or IPS to SS can be observed when changing the coupling strength $\lambda$ and the phase shift $\alpha$.
\begin{figure}
  \includegraphics[width=7.3cm,height=4.8cm]{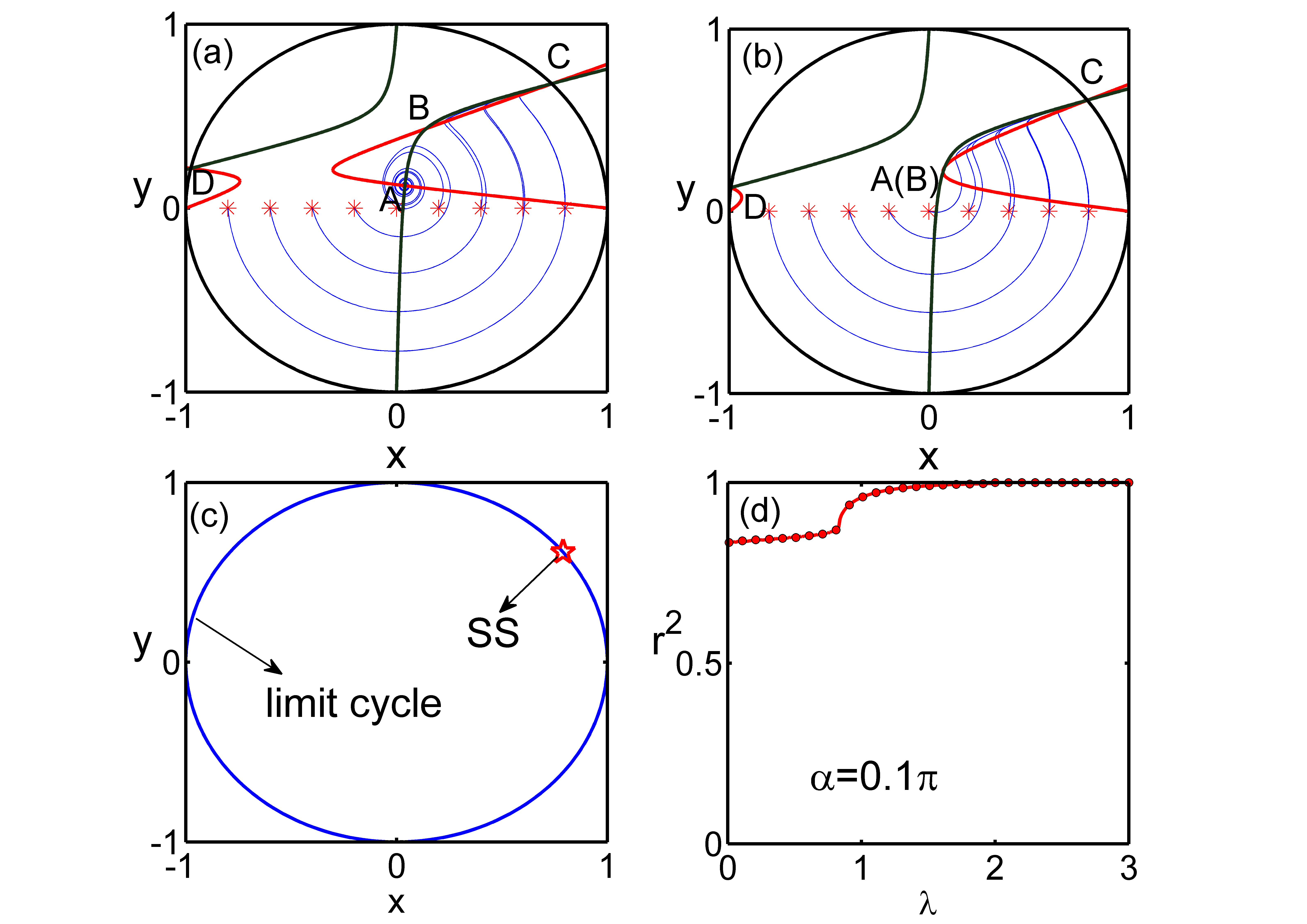}\\
  \caption{(color online) The EOP phase plane for $\Delta\omega=9$, $K=10$, $\alpha=-0.1\pi$, (a) $\lambda=1.8$, (b) $\lambda=2.17$. Red lines are $\dot x=0$, and black $\dot y=0$. The intersections of them are fixed points A-D. Different initial values with trajectories are marked by '$\ast$'. (c) The EOP phase space for $0<\alpha\leq0.5\pi$, the limit circle corresponding to the SPS, and fixed point corresponding to SS. (d) The order parameter against the corresponding coupling strength.}\label{fig:3}
\end{figure}

In Figs. \ref{fig:3}(a) and \ref{fig:3}(b), we show dynamical manifestations of the discontinuous transition from SPS to SS. Fig. \ref{fig:3}(a) exhibits the coexistence of SPS and SS as the stable fixed points A and C respectively. The basins of attraction of SPS and SS are separated by the repeller B. When $\lambda$ increases, as shown in Fig. \ref{fig:3}(b), the repeller B and the attractor A collide and disappear via an inverse saddle-node bifurcation, and this discontinuous transition makes SS a global attractor.

The abrupt transition implies that there are two critical coupling strengths $\lambda_{c}^{b}$ and $\lambda_{c}^{f}$, where $\lambda_{c}^{b}=\lambda_{sc}^{-}$ and $\lambda_{c}^{f}$ depends on the basion of attraction. The upper limit of $\lambda_{c}^{f}$ can be determined by analyzing the inverse saddle-node bifurcation as
\begin{equation}
\small{\hat\lambda_{c}^{f}=\frac{\Delta\omega}{\sqrt{2K\cos2\alpha+1}}}.
\end{equation}
The bistable/coexistence regime corresponding to the above discontinuous transition is given for $\alpha \in(\alpha_{0}^{-},0]$ and $\lambda \in[\lambda_{sc},\hat\lambda_{c}^{f})$.

Figs. \ref{fig:3}(c) and (d) show the transition from IPS to SS. As shown in Fig. \ref{fig:3}(c) IPS is a limit cycle and SS is a fixed point on this circle. The transition from IPS to SS takes place continuously through a saddle-node bifurcation when the coupling strength is increased, as shown in Fig. \ref{fig:3}(d).

\begin{figure}
  \includegraphics[width=7.5cm,height=4.8cm]{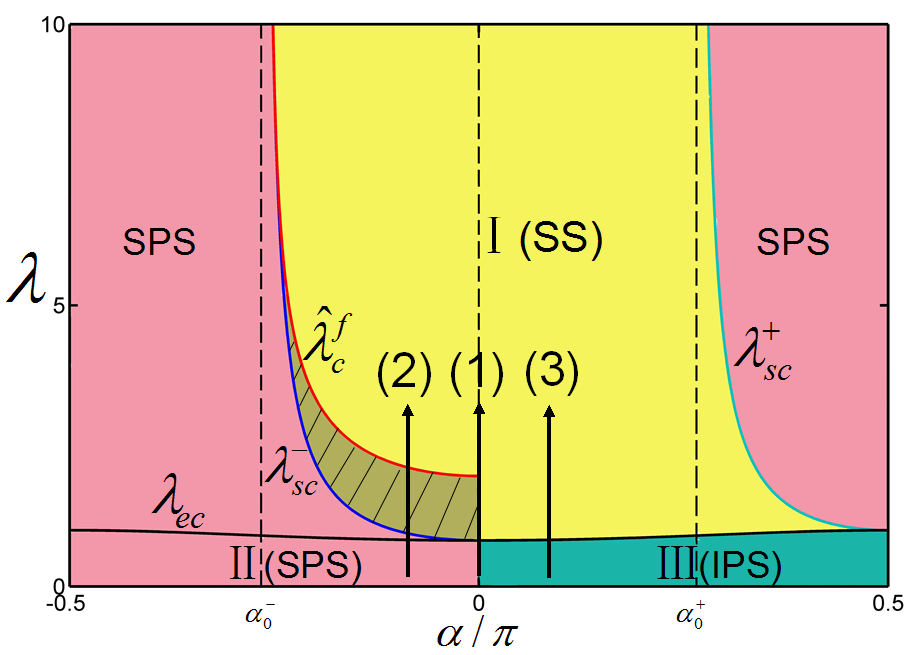}\\
  \caption{(color online) Phase diagram of system Eq. (\ref{equ:5}). Regime $\textrm{\uppercase\expandafter{\romannumeral1}}$ is stable synchronization region. Regimes $\textrm{\uppercase\expandafter{\romannumeral2}}$ and $\textrm{\uppercase\expandafter{\romannumeral3}}$ are asynchronous region with different incoherent, SPS and IPS respectively. The coexistence regime of incoherent state and SS is plotted by shadow. Three routes to synchronization are shown as SPS to SS, IPS to SS, and neutral state to SS.}
  \label{fig:2}
\end{figure}

A phase diagram describing various dynamical states and transitions of system Eqs. (\ref{equ:4}) is given in Fig. \ref{fig:2}, where regime $\textrm{\uppercase\expandafter{\romannumeral1}}$ is the stable region for SS and regime $\textrm{\uppercase\expandafter{\romannumeral2}}$ for SPS, regime $\textrm{\uppercase\expandafter{\romannumeral3}}$ for IPS. Three routes from incoherent state to SS are shown as (1) neutral state to SS, (2) SPS to SS, and (3) IPS to SS. The structure relationship of incoherent state and SS determines the feature of the transition.

A scale-free (SF) network with a low mean degree can be considered as a collection of star graphs \cite{7,19,25}. The above results can be applied straightly to studies of ES in SF network. Using Barabasi-Albert model with $m_{0}=1$ \cite{21} as an example, we generate a SF network with $N=500$ nodes and $K=26$ nodes in the largest star motif.

In Figs. \ref{fig:5}(a)-(c), the order parameters of the SF network $r$ and the largest star motif $r_{L}$ for three different routes to SS are given. The abrupt transition from neutral state to SS and SPS to SS are shown in Figs. \ref{fig:5}(a) and \ref{fig:5}(b) respectively, and the continuous transition from IPS to SS is shown in Fig. \ref{fig:5}(c). It is clear that the largest star motif and the SF network share the same properties of synchronous behaviors, such as the type of transition, either abrupt or continues, and the same critical coupling strengthes. Therefore the synchronization transition of SF networks can be well understood in terms of the above discussions on star networks. The original explosive synchronization(ES) of SF network corresponds to the path from neutral state to SS, and the property of neutral state is checked for the largest star motif in SF network in Fig. \ref{fig:5}(d), where the order parameters depend on initial conditions.

\begin{figure}
  \includegraphics[width=7.5cm,height=4.8cm]{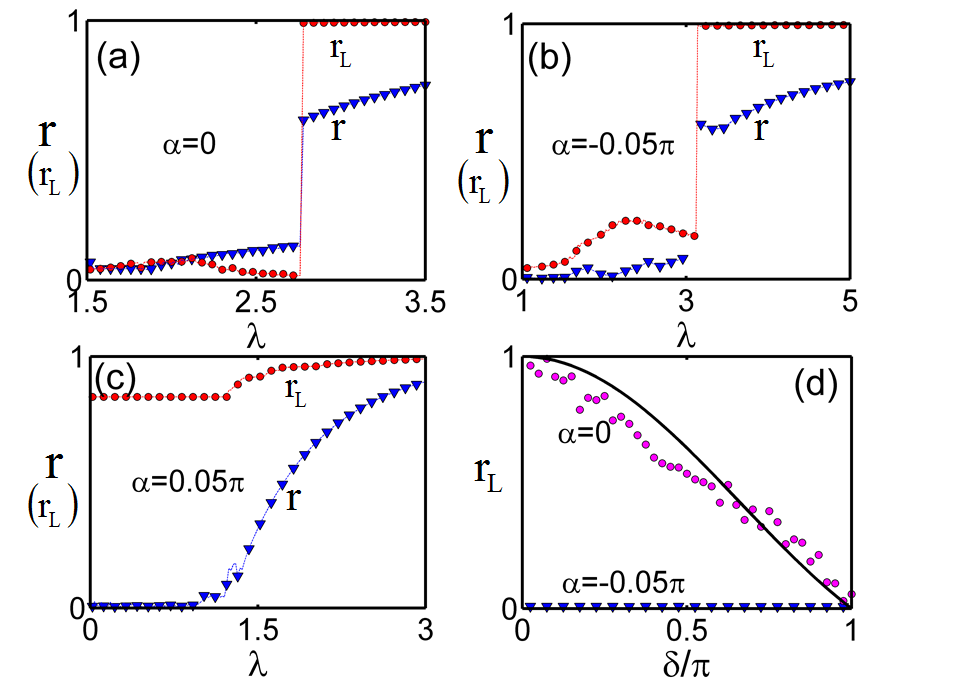}\\
  \caption{(color online) (a)-(c): The order parameters of the SF network $r$ and the largest star motif $r_{L}$ varies with the coupling strength for different $\alpha$. (d) $r_{L}$ with different initial states randomly chosen from $[-\delta,\delta]$ with different $\alpha$, and $\lambda=0.3$, the black solid line is theoretical initial order parameter $r=\sin \delta / \delta$. The size of network is $N=500$.}\label{fig:5}
\end{figure}

To summarize, in this Letter we proposed the EOP equation in terms of the OA approach to study the synchronization of coupled oscillators on a star graph. By reducing from a high-dimensional phase space  to a much lower-dimensional order parameter space without additional approximation, one is able to grasp analytically the essential dynamical mechanism of different scenarios of synchronization. Different solutions of the EOP equation such as fixed points and limit cycles build correspondences with different collective states of coupled oscillators, and different bifurcations reveal various transitions among collective states. In the bistable regime, the measure of synchronized and the incoherent states can be analytically obtained by using the EOP dynamics, which is a very sophisticated and analytically inaccessible procedure in the phase space of coupled oscillators. The analysis and results in the present work can be naturally applicable to scale-free networks, where the star topology plays a dominant role in governing collective dynamics. The properties of three routes to synchronization proposed in star networks are also shown in SF networks, which pave the way for analyzing the relation between star motif and SF network and help us understand the transition to synchronization in more general heterogenous networks.

This work is partially supported by the National Natural Science Foundation of China (Grant No. 11075016 and 11475022).\\


\begin{thebibliography}{99}
\bibitem{1} Y. Kuramoto, \textit{Chemical Oscillations, Waves and Turbulence}(Springer, Berlin, 1984).
\bibitem{2} J.A. Acebron, L.L. Bonilla, C.J.P. Vicente, F. Ritort, and R. Spigler, Rev. Mod. Phys. \textbf{77}, 137 (2005).
\bibitem{3} S.H. Strogatz, Physica D \textbf{143}, 1 (2000).
\bibitem{4} A. Pikovsky, M. Rosenblum, and J. Kurths, \textit{Synchronization:a Universal Concept in Nonlinear Sciences} (Cambridge University Press, Cambridge, England, 2001).
\bibitem{5} S.N. Dorogovtsev, A.V. Goltsev, and J.F.F. Mendes, Rev. Mod.Phys. \textbf{80}, 1275 (2008).
\bibitem{6} Arenas A, Diaz-Guilera A, Kurths J, Moreno Y, and Zhou C, Phys. Rep. \textbf{469} (2008) 93.
\bibitem{6'}Z. Zheng, G. Hu, and B. Hu, Phys. Rev. Lett. \textbf{81}, 5318 (1998).
\bibitem{7} J. Gomez-Gardenes, S. Gomez, A. Arenas, and Y. Moreno, Phys. Rev. Lett. \textbf{106}, 128701 (2011).
\bibitem{8} ThomasKaueDalMaso. Peron, and  F.A. Rodrigues, Phys. Rev. E \textbf{86}, 016102(2012).
\bibitem{9}P. Ji, T.K.DM. Peron,  P.J. Menck, F.A. Rodrigues, and J. Kurths, Phys. Rev. Lett. \textbf{110}, 218701 (2013).
 \bibitem{10}I. Leyva, R. Sevilla-Escoboza, J.M. Buldu, I. Sendina-Nadal, J. Gomez-Gardenes, A. Arenas, Y. Moreno, S. Gomez, R. Jaimes-Reategui, and S. Boccaletti, Phys. Rev. Lett. \textbf{108},168702 (2012).
 \bibitem{11}X. Zhang, X. Hu, J. Kurths, and Z. Liu, Phys. Rev. E \textbf{88}, 010802(R) (2013).
  \bibitem{12}P. Li, K. Zhang, X. Xu, J. Zhang, and M. Small, Phys. Rev.
E.\textbf{87}, 042803 (2013).
 \bibitem{13} I. Leyva, A. Navas, I. Sendina-Nadal, J.A. Almendral, J. M. Buldu, M. Zanin, D. Papo, and S. Boccaletti, Sci. Rep. 3. 1281(2013).
 \bibitem{13'}X. Hu, S. Boccaletti, W. Huang, X. Zhang, Z. Liu, S. Guan, and Choy-Heng Lai Sci. Rep. 4. 7262(2014).
 \bibitem{14} H. Sakaguchi and Y. Kuramoto, Prog. Theor. Phys. \textbf{76}, 576(1986).
 \bibitem{15} O.E. Omel'chenko and M. Wolfrum, Phys. Rev. Lett. \textbf{109}, 164101(2012).
 \bibitem{16} S. Watanabe, S.H. Strogatz, Physica D \textbf{74} 197-253(1994).
 \bibitem{17} E. Ott and T.M. Antonsen, Chaos \textbf{18}, 037113(2008).
 \bibitem{18} S.A. Marvel and S.H. Strogatz, Chaos \textbf{19}, 013132(2009).
 \bibitem{19} Y. Zou, T. Pereira, M. Small, Z. Liu, and J. Kurths, Phys. Rev. Lett. \textbf{112}, 114102(2014).
 \bibitem{20} J. Gao, Y. Sun, C. Xu, and Z. Zheng, (to be submitted).
 \bibitem{21} R. Albert and A. Barabasi, Rev. Mod. Phys.\textbf{74},
     47(2002).
 \bibitem{22} S.A. Marvel, R.E. Mirollo, and S.H. Strogatz, Chaos \textbf{19}, 043104 (2009).
 \bibitem{23} D. Topaj and A. Pikovsky, Physica D \textbf{170}, 118 (2002).
 \bibitem{24} S. Watanabe and S. H. Strogatz, Phys. Rev. Lett. \textbf{70}, 2391 (1993).
 \bibitem{25} K. Judd, M. Small, and T. Stemler, Europhys. Lett. \textbf{103}, 58004 (2013).
\end{thebibliography}
\end{document}